\documentclass[10pt,letterpaper]{article} 

\usepackage{amsmath,amsfonts,amsthm,amssymb,stmaryrd,graphicx,relsize,bm}

\theoremstyle{plain}

\numberwithin{equation}{section}

\newenvironment{exam}[1]
{\begin{flushleft}\textbf{Example #1}.\enspace}%
{\end{flushleft}}

\newcommand{\integers}{{\mathbb Z}}
\newcommand{\positive}{{\mathbb N}}
\newcommand{\hscript}{{\mathcal H}}
\newcommand{\sscript}{{\mathcal S}}
\newcommand{\sscripthat}{\widehat{\sscript}}
\newcommand{\bfe}{\mathbf{e}}
\newcommand{\bff}{\mathbf{f}} 
\newcommand{\bfg}{\mathbf{g}} 
\newcommand{\bfp}{\mathbf{p}} 

\newcommand{\bfx}{\mathbf{x}}
\newcommand{\bfy}{\mathbf{y}} 
\newcommand{\bfzero}{\mathbf{0}}
\newcommand{\ctimes}{\mathrel{\mathlarger\cdot}}
  
\newcommand{\ab}[1]{\left|#1\right|}
\newcommand{\doubleab}[1]{\left|\left|#1\right|\right|}
\newcommand{\brac}[1]{\left\{#1\right\}}
\newcommand{\paren}[1]{\left(#1\right)}
\newcommand{\sqbrac}[1]{\left[#1\right]}
\newcommand{\elbows}[1]{{\left\langle#1\right\rangle}}
\newcommand{\ket}[1]{{\left|#1\right>}}
\newcommand{\bra}[1]{{\left<#1\right|}}

\begin{document}

\title{RECONDITIONING\\IN\\DISCRETE QUANTUM FIELD THEORY
}
\author{S. Gudder\\ Department of Mathematics\\
University of Denver\\ Denver, Colorado 80208, U.S.A.\\
sgudder@du.edu
}
\date{}
\maketitle

\begin{abstract}
We consider a discrete scalar, quantum field theory based on a cubic 4-dimensional lattice. We mainly investigate a discrete scattering operator $S(x_0,r)$ where $x_0$ and $r$ are positive integers representing time and maximal total energy, respectively. The operator $S(x_0,r)$ is used to define transition amplitudes which are then employed to compute transition probabilities. These probabilities are conditioned on the time-energy $(x_0,r)$. In order to maintain total unit probability, the transition probabilities need to be reconditioned at each $(x_0,r)$. This is roughly analogous to renormalization in standard quantum field theory, except no infinities or singularities are involved. We illustrate this theory with a simple scattering experiment involving a common interaction Hamiltonian. We briefly mention how discreteness of spacetime might be tested astronomically. Moreover, these tests may explain the existence of dark energy and dark matter.
\end{abstract}

\section{Introduction}  
In a recent paper, the author discussed a version of discrete quantum field theory \cite{gud16}. In the present article we present some changes and refinements of these ideas. One of our main refinements is the introduction of reconditioning for transition probabilities. As in the previous work, we neglect spin and only consider scalar quantum fields.

The paper begins with a description of discrete spacetime $\sscript$ on a cubic 4-dimensional lattice. The dual space $\sscripthat$ describes a discrete energy-momentum structure. Because of the discreteness of $\sscripthat$, particles with a given mass and total energy can only propagate in a finite number of possible directions. We point out in the last section that this may provide an explanation for dark energy and dark matter. It may also lead to astronomical methods of experimentally verifying the discreteness of spacetime.

We begin by defining free quantum fields $\phi (x,r)$, $x\in\sscript$, $r\in\positive$ for particles of mass $m$. The $x\in\sscript$ represents a spacetime point and $r\in\positive$ is a maximal total energy. The advantage of considering $\phi (x,r)$ is that it can be written in closed form as a finite sum. We briefly consider the commutant
$\sqbrac{\phi (x,r),\phi (y,r)}$ and mention its relationship to quantum locality. Our main investigation concerns a discrete scattering operator $S(x_0,r)$ where $x_0,r\in\positive$ represent time and maximal total energy, respectively. The operator $S(x_0,r)$ is used to define transition amplitudes which are then employed to compute transition probabilities conditioned on the time-energy $(x_0,r)$. In order to maintain total unit probability, the transition probabilities need to be reconditioned at each
$(x_0,r)$. This is roughly analogous to renormalization in standard quantum field theory \cite{ps95,vel94} except no infinities or singularities are involved. The operators $S(x_0,r)$ are not unitary in general, and we do not know whether
\begin{equation*}
\lim _{x_0,r\to\infty}S(x_0,r)
\end{equation*}
is unitary or even exists in some sense. Fortunately, this existence is not needed for approximations and all that is necessary is the reconditioning procedure.

We illustrate this theory with a simple two-particle scattering experiment involving a standard interaction Hamiltonian. It is interesting that even in the first nontrivial perturbation presented here, the computations get rather involved and take many steps. This indicates that computer assisted calculations will be necessary to obtain accurate results from higher order perturbations.

\section{Discrete Spacetime and Energy-Momentum}  
Our basic assumption is that spacetime is discrete and has the form of a 4-dimensional cubic lattice $\sscript$ \cite{bdp16,cro16,gud16}. We then have
$\sscript =\integers ^+\times\integers ^3$ where $\integers ^+=\brac{0,1,2,\ldots}$ represents discrete time and $\integers =\brac{0,\pm 1,\pm 2,\ldots}$ so
$\integers ^3$ represents discrete 3-space. If $x=(x_0,x_1,x_3,x_3)\in\sscript$, we sometimes write $x=(x_0,\bfx )$ where $x_0\in\integers ^+$ is time and
$\bfx\in\integers ^3$ is a 3-space point. We equip $\sscript$ with the Minkowski distance
\begin{equation*}
\doubleab{x}_4^2=x_0^2-\doubleab{\bfx}_3^2=x_0^2-x_1^2-x_2^2-x_3^2
\end{equation*}
As usual $x,y\in\sscript$ are \textit{time-like separated} if $\doubleab{x-y}_4^2\ge 0$ and \textit{space-like separated} if $\doubleab{x-y}_4<0$. Of course, we are using units in which we take the speed of light to be 1. The vectors
\begin{align*}
d&=(1,\bfzero )=(1,0,0,0)\\
e&=(0,\bfe )=(0,1,0,0)\\
f&=(0,\bff )=(0,0,1,0)\\
g&=(0,\bfg )=(0,0,0,1)\\
\end{align*}
give a basis for $\sscript$ in the sense that every $x\in\sscript$ has a unique form
\begin{equation*}
x=nd+me+pf+qg,\quad n\in\positive ^+, m,p,q\in\positive
\end{equation*}

The dual of $\sscript$ is denoted by $\sscripthat$. We regard $\sscripthat$ as having the identical structure as $\sscript$ and that $\sscripthat$ again has basis $d,e,f,g$. The only difference is that we denote elements of $\sscript$ by
\begin{equation*}
p=(p_0,\bfp )=(p_0,p_1,p_2,p_3)
\end{equation*}
and interpret $p$ as the energy-momentum vector for a particle. In fact, we sometimes call $p\in\sscript$ a particle. Moreover, we only consider the forward cone
\begin{equation*}
\brac{p\in\sscripthat\colon\doubleab{p}_4\ge 0}
\end{equation*}
For a particle $p\in\sscript$ we call $p_0\ge 0$ the \textit{total energy}, $\doubleab{\bfp}_3\ge 0$ the \textit{kinetic energy} and $m=\doubleab{p}_4$ the \textit{mass} of $p$. The integers $p_1,p_2,p_3$ are \textit{momentum components}. Since
\begin{equation*}
m^2=\doubleab{p}_4^2=p_0^2-\doubleab{\bfp}_3^2
\end{equation*}
we conclude that Einstein's energy formula
\begin{equation*}
p_0=\sqrt{m^2+\doubleab{\bfp}_3^2}
\end{equation*}
holds.

We now consider some examples. Suppose that $m=\sqrt{3}$ and $p_0=2$. Then
\begin{equation*}
p_1^2+p_2^2+p_3^2=\doubleab{\bfp}_3^2=p_0^2-m^2=1
\end{equation*}
It follows that the momentum $\bfp$ is one of the following six vectors $\bfp =\pm\bfe ,\ \pm\bff ,\ \pm\bfg$. Now suppose that $m=\sqrt{3}$ and $p_0=3$. Now there are 24 possible momentum vectors
\begin{equation*}
\bfp =\pm 2\bfe\pm\bff\pm\bfg ,\ \pm 2\bff\pm\bfe\pm\bfg ,\ \pm 2\bfg\pm\bfe\pm\bff
\end{equation*}

Tables for many more examples are given in \cite{gud16} but for this work we shall only require the photon momentum vectors given in Table~1. For a photon we have $m=0$ and $p_0=\doubleab{\bfp}_3$.

\vglue 1pc
\begin{table}
\scalebox{0.9}{
\parindent=-3pc
\begin{tabular}{c|c|c}
&&Number of\hfill\hfill\\
$p_0$&Momentum Vectors&Vectors\hfill\hfill\\
\hline
1&$\pm\bfe ,\ \pm\bff ,\ \pm\bfg$\hfill\hfill&6\\\hline
2&$\pm 2\bfe ,\ \pm 2\bff ,\ \pm 2\bfg$\hfill\hfill&6\\\hline
3&$\pm 3\bfe ,\ \pm 3\bff ,\ \pm 3\bfg ,\ \pm 2\bfe\pm 2\bff\pm\bfg ,\ \pm 2\bfe\pm 2\bfg\pm\bff ,\ \pm 2\bff\pm 2\bfg\pm\bfe$\hfill\hfill&30\\\hline
4&$\pm 4\bfe ,\ \pm 4\bff ,\ \pm 4\bfg$\hfill\hfill&6\\\hline
5&$\pm 5\bfe ,\ \pm 5\bff ,\ \pm 5\bfg ,\ \pm 4\bfe\pm 3\bff ,\ \pm 4\bfe\pm 3\bfg ,\ \pm 4\bff\pm 3\bfe$,\hfill\hfill&30\\
&$\pm 4\bff\pm 3\bfg ,\ \pm 4\bfg\pm3\bfe ,\ \pm 4\bfg\pm 3\bff$\hfill\hfill&\\\hline
6&$\pm 6\bfe ,\ \pm 6\bff ,\ \pm 6\bfg ,\ \pm 4\bfe\pm 4\bff\pm 2\bfg ,\ \pm 4\bfe\pm 4\bfg\pm 2\bff ,\ \pm 4\bff\pm 4\bfg\pm 2\bfe$\hfill\hfill&30\\\hline
7&$\pm 7\bfe ,\ \pm 4\bff ,\ \pm 4\bfg ,\ \pm 6\bfe\pm 3\bff\pm 2\bfg ,\ \pm 6\bff\pm 3\bfg\pm 2\bfe ,\ \pm 6\bfg\pm 3\bfe\pm 2\bff$\hfill\hfill&\\
&$\pm 6\bfe\pm 3\bfg\pm 2\bff ,\ \pm 6\bff\pm 3\bfe\pm 2\bfg ,\ \pm 6\bfg\pm 3\bff\pm 2\bfe$\hfill\hfill&54\\\hline
8&$\pm 8\bfe ,\ \pm 8\bff ,\ \pm 8\bfg$\hfill\hfill&30\\\hline
9&$\pm 9\bfe ,\ \pm 9\bff ,\ \pm 9\bfg ,\ \pm 7\bfe\pm 4\bff\pm 4\bfg ,\ \pm 7\bff\pm 4\bfg\pm 4\bfe ,\ \pm 7\bfg\pm 4\bfe\pm 4\bff$\hfill\hfill&102\\
&$\pm 6\bfe\pm 6\bff\pm 3\bfg ,\ \pm 6\bfe\pm 6\bfg\pm 3\bff ,\ \pm 6\bff\pm 6\bfg\pm 3\bfe , \pm 8\bfe\pm 4\bff\pm\bfg$\hfill\hfill&\\
&$\pm 8\bff\pm 4\bfg\pm\bfe ,\ \pm 8\bfg\pm 4\bfe\pm\bff ,\ \pm 8\bfe\pm 4\bfg\pm\bff ,\ \pm 8\bff\pm 4\bfe\pm\bfg$\hfill\hfill&\\
&$\pm 8\bfg\pm 4\bff\pm\bfe$\hfill\hfill&\\\hline
10&$\pm 10\bfe ,\ \pm 10\bff ,\ \pm 10\bfg ,\ \pm 8\bfe\pm 6\bff ,\ \pm 8\bfe\pm 6\bfg ,\ \pm 8\bff\pm 6\bfg$\hfill\hfill&30\\
&$\pm 8\bff\pm 6\bfe ,\ \pm 8\bfg\pm 6\bff ,\ \pm 8\bfg\pm 6\bfe$\hfill\hfill&\\
\hline\noalign{\medskip}
\multicolumn{3}{c}{\textbf{Table 1}}\\
\end{tabular}}
\end{table}
\parindent=18pt

The number of momentum vectors can easily get into the thousands even for moderate size $p_0$. For example, with $p_0=91$ there are 702 different momentum vectors (the author is indebted to Joel Cohen for this example). For brevity, we list them without regard to order and $\pm$ signs:
\begin{align*}
&\brac{91},\brac{35,84},\brac{6,18,89},\brac{6,26,87},\brac{6,39,82},\brac{6,59,73},\\
&\brac{9,10,90},\brac{9,46,78},\brac{9,62,66}\brac{18,54,71},\brac{21,28,89},\\
&\brac{26,39,78},\brac{26,57,66},\brac{30,55,56}\brac{39,54,62},\brac{46,54,57}
\end{align*}

\section{Free Quantum Fields} 
We assume that we are describing a physical system that contains particles that can be a finite number of various types. For illustrative purposes, suppose there are two types of particles under consideration that we call $p$-particles and $q$-particles. To describe the particles quantum mechanically, we first construct a complex Hilbert space $K$. Technically speaking, $K$ is a symmetric Fock space, but the details are not important here. All we need to know is that $K$ exists and has a very descriptive orthonormal basis of the form
\begin{equation*}
\ket{p^1p^2\cdots p^nq^1q^2\cdots q^s}
\end{equation*}
that represents the quantum state in which there are $n$ $p$-particles and $s$ $q$-particles where $p^i,q^j\in\sscripthat$, $i=1,\ldots ,n$, $j=1,\ldots ,s$ \cite{gud16}. The \textit{vacuum} state in which there are no particles present is the unit vector $\ket{0}$. If there are more than two types of particles present, we can proceed analogously.

For a $p$-particle $p\in\sscripthat$, we define the \textit{annihilation operator} $a(p)$ on $K$ by $a(p)\ket{0}=0$ where 0 is the zero vector and
\begin{equation*}
a(p)\ket{pp\cdots pp^1\cdots p^nq^1\cdots q^s}=\sqrt{n}\,\ket{pp\cdots pp^1\cdots p^nq^1\cdots q§ ^s}
\end{equation*}
where $n$ $p$'s appear on the left side and $n-1$ $p$'s appear on the right side. We interpret $a(p)$ as the operator that annihilates a particle with energy-momentum $p$. We also have the annihilation operator $a(q)$ that acts analogously for $q$-particles. The adjoint $a(p)^*$ of $a(p)$ is the operator on $K$ defined by
\begin{equation*}
a(p)^*\ket{pp\cdots pp^1\cdots p^nq^1\cdots q^s}=\sqrt{n+1}\,\ket{pp\cdots pp^1\cdots p^nq^1\cdots q^s}
\end{equation*}
where $n$ $p$'s appear on the left side and $n+1$ $p$'s appear on the right side. We call $a(p)^*$ the \textit{creation operator} on $K$. We interpret $a(p)^*$ as the operator that creates a particle with energy-momentum $p$. The operators $a(p)$ and $a(p)^*$ have the characteristic property that their commutator
$\sqbrac{a(p),a(p)^*}=I$.

Suppose $p$-particles have mass $m$. We define the \textit{mass hyperboloid}
\begin{equation*}
\Gamma _m=\brac{p\in\sscripthat\colon\doubleab{p}_4=m}
\end{equation*}
Moreover, for $x\in\sscript$, $p\in\sscripthat$ we define the indefinite inner product
\begin{equation*}
px=p_0x_0-p_1x_1-p_2x_2-p_3x_3
\end{equation*}
For any $x\in\sscript$, $r\in\positive$, we define the \textit{free quantum field} at $x$ with mass $m$ and maximal total energy $r$ by
\begin{equation}         
\label{eq31}
\phi (x,r)=\sum\brac{\frac{1}{p_0}\,\sqbrac{a(p)e^{i\pi px/2}+a(p)^*e^{-i\pi px/2}}\colon p\in\Gamma _m,p_0\le r}
\end{equation}
Notice that $\phi (x,r)$ is given as a closed form expression in terms of a finite summation.

It follows from the work in \cite{gud16} that for any free quantum field $\phi$ the commutator
\begin{align}         
\label{eq32}
&\sqbrac{\phi (x,r),\phi (y,r)}\\
&\quad =2i\sum\brac{\frac{1}{p_0^2}\,\sin\sqbrac{\pi p_0(x_0-y_0)/2}\cos\sqbrac{\pi\bfp\ctimes (\bfx -\bfy )/2}\colon p\in\Gamma _m,p_0\le r}I\notag
\end{align}
As a consequence of \eqref{eq32} we have the \textit{equal time commutation relation}: if $x_0=y_0$, then $\sqbrac{\phi (x,r),\phi (y,r)}=0$ for every $r\in\positive$. Moreover, from \eqref{eq32} we have that
\begin{equation*}
\lim _{r\to\infty}\sqbrac{\phi (x,r),\phi (y,r)}
\end{equation*}
exists for all $x,y\in\sscript$. Letting $\phi (x)=\lim\limits _{r\to\infty}\phi (x,r)$, \textit{quantum locality} states that if $\doubleab{x-y}_4^2<0$ then $\sqbrac{\phi (x),\phi (y)}=0$. That is, if $x$ and $y$ are space-like separated, then a field measurement at $x$ does not affect a field measurement at $y$. If $x_0=y_0$ and $x\ne y$, then
$\doubleab{x-y}_4<0$ so in this case quantum locality holds. Presumably, quantum local holds in standard quantum field theory \cite{ps95,vel94}. However, we now give two examples which show that quantum locality holds for certain space-like separated pairs and probably does not hold for others in this discrete theory.
\vglue 2pc

\begin{exam}{1} 
If $x=(1,1,1,1)$ and $y=0$, then $\doubleab{x-y}_4^2=-2<0$ Letting $m=0$ (photon), applying \eqref{eq31} we have that
\begin{align*}
&\sqbrac{\phi (x,r),\phi (y,r)}\\
&\quad =2i\sum\brac{\frac{1}{p_0^2}\,\sin (\pi p_0/2)\cos\sqbrac{\pi (p_1+p_2+p_3)/2}\colon p\in\Gamma _0,p_0\le r}I
\end{align*}
Now if $p_0$ is even, then $\sin (\pi p_0/2)=0$ so these terms vanish in the summation. Otherwise, $p_0$ is odd. Since
\begin{equation*}
p_0^2=p_1^2+p_2^2+p_3^2
\end{equation*}
it follows that $p_1$, $p_2$ and $p_3$ are all odd or two of these numbers are even and one is odd. In either case
\begin{equation*}
\cos\sqbrac{\pi (p_1+p_2+p_3)/2}=0
\end{equation*}
so these terms vanish in the summation. We conclude that $\sqbrac{\phi (x,r),\phi (y,r)}=0$ for all $r\in\positive$ so quantum locality holds in this case.
\quad\qedsymbol
\end{exam} 
\vglue 1pc

\begin{exam}{2} 
If $x=(1,2\bfe )$ and $y=0$, then $\doubleab{x-y}_4^2=-3<0$. Again, letting $m=0$ and applying \eqref{eq31} we have that
\begin{equation*}
\sqbrac{\phi (x,r),\phi (y,r)}=2i\sum\brac{\frac{1}{p_0^2}\,\sin (\pi p_0/2)\cos (\pi p_1)\colon p\in\Gamma _0,p_0\le r}I
\end{equation*}
From Table~1 we obtain
\begin{align*}
\sqbrac{\phi (x,1),\phi (y,1)}&=\sqbrac{\phi (x,2),\phi (y,r)}=4iI\\
\sqbrac{\phi (x,3),\phi (y,3)}&=\sqbrac{\phi (x,4),\phi (y,4)}=4iI-\frac{2i}{9}\, (10)I=1.777iI\\
\sqbrac{\phi (x,5),\phi (y,5)}&=\sqbrac{\phi (x,6),\phi (y,6)}=1.777iI+\frac{2i}{25}\,(10)I=2.577iI\\\noalign{\medskip}
\sqbrac{\phi (x,7),\phi (y,7)}&=\sqbrac{\phi (x,8),\phi (y,8)}=2.577iI-\frac{2i}{49}\,(18)I=1.843iI\\\noalign{\medskip}
\sqbrac{\phi (x,9),\phi (y,9)}&=\sqbrac{\phi (x,10),\phi (y,10)}=1.843iI+\frac{2i}{81}\,(34)I=2.683iI
\end{align*}
We conjecture that $\lim\limits _{r\to\infty}\sqbrac{\phi (x,r),\phi (y,r)}\ne 0$ so quantum locality does not hold in this case.\ \qedsymbol
\end{exam} 

\section{Scattering Operators} 
An \textit{interaction Hamiltonian} is a set of self-adjoint operators $H(x_0,r)$, $x_0\in\positive\cup\brac{0}$, $r\in\positive$ on $K$ where, as before $x_0$ represents time and
$r$ represents a maximal total energy. The corresponding scattering operators $S(x_0,r)$ satisfy the initial condition $S(0,r)=I$ and the ``second quantization'' condition
\begin{equation}         
\label{eq41}
\nabla _{x_0}S(x_0,r)=iH(x_0,r)S(x_0,r)
\end{equation}
for all $r\in\positive$ where $\nabla _{x_0}$ is the difference operator given by
\begin{equation*}
\nabla _{x_0}S(x_0,r)=S(x_0+1,r)-S(x_0,r)
\end{equation*}
It is easy to show \cite{gud16} that from \eqref{eq41} we have that
\begin{align}         
\label{eq42}
S(x_0,r)&=\sqbrac{I+iH(x_0-1,r)}\sqbrac{I+iH(x_0-2,r)}\cdots\sqbrac{I+iH(0,r)}\notag\\
  &=I+i\sum _{j=0}^{n-1}H(j,r)+i^2\sum _{j_2<j_1}^{n-1}H(j_1,r)H(j_2,r)\\
  &\qquad +i^3\sum _{j_3<j_2<j_1}H(j_1,r)H(j_2,r)H(j_3,r)\notag\\
  &\qquad +\cdots +i^nH(n-1,r)H(n-2,r)\cdots H(0,r)\notag
\end{align}
In general, the operators $H(x_0,r)$ do not commute and the order in \eqref{eq42} must be maintained. This is called the \textit{time-ordered product} of operators. For
$x_0\ne 0$, $S(x_0,r)$ is not unitary in general. However, this problem is circumvented by the reconditioning procedure that we now discuss.

Reconditioning is roughly related to renormalization. One can give a general definition of reconditioning, but for simplicity we shall illustrate this procedure with an example. Suppose we are interested in a two particle scattering interaction $\ket{pq}\to\ket{p'q'}$. It is shown in \cite{gud16} that corresponding to the incoming (or input) state $\ket{pq}$ there are finitely many possible outgoing (or output) states $\ket{p^jq^j}$, $j=1,\dots ,n$. Let $S(x_0,r)$, $x_0,r\in\positive$ be the scattering operators for this interaction. Then
\begin{equation*}
\bra{p^jq^j}S(x_0,r)\ket{pq}
\end{equation*}
$j=1,\ldots ,n$ are the conditional amplitudes at the perturbation level $(x_0,r)$ given the input state $\ket{pq}$. The \textit{probability} at $(x_0,r)$ of the interaction
$\ket{pq}\to\ket{p^jq^j}$ is defined by
\begin{equation}         
\label{eq43}
P_{x_0,r}\paren{\ket{pq}\to\ket{p^jq^j}}=\frac{\ab{\bra{p^jq^j}S(x_0,r)\ket{pq}}^2}{\sum\limits _{j=1}^n\ab{\bra{p^jq^j}S(x_0,r)\ket{pq}}^2}
\end{equation}
The denominator in \eqref{eq43} is needed because it is not unity in general, so a renormalization (reconditioning) is needed at each perturbation level $(x_0,r)$. Conditions must be placed on the interaction Hamiltonian $H(x_0,r)$ so that
\begin{equation*}
P\paren{\ket{pq}\to\ket{p^jq^j}}=\lim _{x_o,r\to\infty}P_{x_0,r}\paren{\ket{pq}\to\ket{p^jq^j}}
\end{equation*}
exists for all $j=1,\ldots ,n$. In this case the reconditioning is successful and $P\paren{\ket{pq}\to\ket{p^jq^j}}$ is a probability in the sense that
\begin{equation*}
\sum _{j=1}^nP\paren{\ket{pq}\to\ket{p^jq^j}}=1
\end{equation*}
The interaction Hamiltonian $H(x_0,r)$ is usually constructed from a \textit{Hamiltonian density} consisting of self-adjoint operators on $K$ denoted by $\hscript (x,r)$,
$x\in\sscript$, $r\in\positive$. The \textit{space-volume} at $x_0$ is defined by the cardinality
\begin{equation*}
V(x_0)=\ab{\brac{\bfx\colon\doubleab{\bfx}_3\le x_0}}
\end{equation*}
We then define the corresponding interaction Hamiltonian to be
\begin{equation}         
\label{eq44}
H(x_0,r)=\frac{1}{V(x_0)}\,\sum\brac{\hscript\paren{(x_0,\bfx )}\colon\doubleab{\bfx}_3\le x_0}
\end{equation}
The first few terms of \eqref{eq44} are
\begin{align*} 
H(0,r)&=\hscript (0,r)\\
H(1,r)&=\frac{1}{7}\left[\hscript \paren{(1,\bfzero ),r}+\hscript\paren{(1,\bfe ),r}+\hscript\paren{(1,-\bfe ),r}+\hscript\paren{(1,\bff ),r)}\right.\\
      &\qquad \left. +\hscript\paren{(1,-\bff ),r}+\hscript\paren{(1,\bfg ),r}+\hscript\paren{(1,-\bfg ),r}\right]\\
      H(2,4)&=\frac{1}{33}\,\sqbrac{\hscript\paren{(2,\bfzero ),r}+\hscript\paren{(2,\bfe ),r}+\cdots +\hscript\paren{(2,2\bfg ),r}+\hscript\paren{(2,-2\bfg ),r}}
\end{align*}

We now illustrate this theory with a simple example. This example contains the elements that can be generalized to more complicated and realistic situations. Suppose we consider the scattering of two electrons with mass $m$ and initial energy-momentum $p,q\in\sscripthat$. We assume that the electrons interact by exchanging photons and arrive with final energy-momentum $p',q'\in\sscript$. As usual in this article, we neglect spin. Let $\phi$ and $\sigma$ be the quantum fields for the electrons and photons, respectively. Then $\phi$ is given by \eqref{eq31} and
\begin{equation}         
\label{eq45}
\sigma (x,r)=\sum _{k\in\Gamma _0}\brac{\frac{1}{k_0}\,\sqbrac{a(k)e^{i\pi kx/2}+a(k)^*e^{-i\pi k/2}}\colon k\in\Gamma _0,\ k_0\le r}
\end{equation}
The Hamiltonian density is usually constructed by interacting the free quantum fields $\phi$ and $\sigma$. A linear combination of $\phi$ and $\sigma$ will not produce any scattering so we take a simple nonlinear combination of the form
\begin{equation}         
\label{eq46}
\hscript (x,r)=\phi (x,r)^2\sigma (x,r)
\end{equation}
Letting $r=3$ and applying \eqref{eq42} we obtain the scattering operators
\begin{align}         
\label{eq47}
S(0,3)&=I\\    
\label{eq48}   
S(1,3)&=I+iH(0,3)\\    
\label{eq49}   
S(2,3)&=I+i\sqbrac{H(0,3)+H(1,3)}-H(1,3)H(0,3)\\    
\label{eq410}   
S(3,3)&=I+i\sqbrac{H(0,3)+H(1,3)+H(2,3)}-H(1,3)H(0,3)\\
&\quad -H(2,3)H(0,3)-H(2,3)H(1,3)-iH(2,3)H(1,3)H(0,3)\notag
\end{align}
As we shall see, \eqref{eq47} and \eqref{eq48} do not provide for scattering. Equations \eqref{eq49} and \eqref{eq410} are more interesting and describe perturbations at levels (2,3) and (3,3), respectively. Working out \eqref{eq410} becomes quite complicated so for illustrative purposes, we give a concrete example that involves \eqref{eq49} in the next section.

\section{Scattering Example} 
We continue the work on electron-electron scattering of Section~4 with a concrete example. Suppose the electron mass is $m=\sqrt{3}$ and that the input state is $\ket{pq}$ where $p=(2,0,-1,0),q=(3,-1,2,1)\in\sscripthat$ represent two electrons. Under the assumption of conservation of energy-momentum which is a basic axiom of the theory \cite{gud16,ps95,vel94}, it is easy to check that there are only three possibilities:\bigskip

{\obeylines
(a)\enspace $\ket{pq}\to\ket{pq}$ (no scattering)
(b)\enspace $\ket{pq}\to\ket{p'q'}$
(c)\enspace $\ket{pq}\to\ket{p''q''}$}

\noindent where $p'=(2,0,0,-1)$, $q'=(3,-1,1,2)$, $p''=(2,1,0,0)$, $q''=(3,-2,1,1)$.

Applying \eqref{eq47}, the scattering amplitude at perturbation level $(0,3)$ becomes
\begin{equation*}
\bra{pq}S(0,3)\ket{rs}=\elbows{pq\mid rs}
\end{equation*}
We conclude that $\bra{pq}S(0,3)\ket{pq}=1$ and
\begin{equation*}
\bra{pq}S(0,3)\ket{p'q'}=\bra{pq}S(0,3)\ket{p''q''}=0
\end{equation*}
Hence, $P_{0,3}\paren{\ket{pq}\to\ket{pq}}=1$ and
\begin{equation*}
P_{0,3}\paren{\ket{pq}\to\ket{p'q'}}=P_{0,3}\paren{\ket{pq}\to\ket{p''q''}}=0
\end{equation*}
Thus, there is no scattering at this stage. Applying \eqref{eq48}, we have that
\begin{equation*}
\bra{pq}S(1,3)\ket{rs}=\elbows{pq\mid rs}+i\bra{pq}H(0,3)\ket{rs}
\end{equation*}
Because there is a single $\sigma (0,3)$ in $H(0,3)$ we get the same result $\bra{pq}S(1,3)\ket{pq}=1$ and
\begin{equation*}
\bra{pq}S(1,3)\ket{p'q'}=\bra{pq}S(1,3)\ket{p''q''}=0
\end{equation*}
so again, there is no scattering at this stage. Things get interesting at level $(2,3)$. Applying \eqref{eq49} gives 
\begin{align*}
\bra{pq}S(2,3)\ket{rs}&=\elbows{pq\mid rs}+i\sqbrac{\bra{pq}H(0,3)\ket{rs}+\bra{pq}H(1,3)\ket{rs}}\\
  &\quad -\bra{pq}H(1,3)H(0,3)\ket{rs}
\end{align*}
As before, the second term on the right side vanishes and we get
\begin{align}         
\label{eq51}
\bra{pq}S(2,3)\ket{pq}&=1-\bra{pq}H(1,3)H(0,3)\ket{pq}\\
\label{eq52}   
\bra{pq}S(2,3)\ket{p'q'}&=-\bra{pq}H(1,3)H(0,3)\ket{p'q'}\\    
\label{eq53}   
\bra{pq}S(2,3)\ket{p''q''}&=-\bra{pq}H(1,3)H(0,3)\ket{p''q''}
\end{align}
We could continue to level $(3,3)$ using \eqref{eq410} but the situation becomes quite complicated and we shall be content to investigate level $(2,3)$.

We see from \eqref{eq51}, \eqref{eq52}, \eqref{eq53} that we need to study $H(1,3)H(0,3)$. Writing this out, we have that
\begin{align*}
H(1,3)H(0,3)&=\frac{1}{7}\,\left[\phi\paren{(1,\bfzero ),3}^2\phi (0,3)^2\sigma\paren{(1,\bfzero ),3}\sigma (0,3)\right.\\
  &\quad +\phi\paren{(1,\bfe )3}^2\phi (0,3)^2\sigma\paren{(1,\bfe ),3}\sigma (0,3)\\
  &\quad \left.+\cdots +\phi\paren{(1,-\bfg ),3}^2\phi (0,3)^2\sigma\paren{(1,-\bfg ),3}\sigma (0,3)\right]
\end{align*}
Treating the photon fields $\sigma$ first, we have that
\begin{align*}
\sigma\paren{(1,\bfzero ),3}\sigma (0,3)&=\sum\brac{\frac{1}{k_0}\,\sqbrac{a(k)e^{i\pi k_0/2}+a(k)^*e^{-i\pi k_0/2}}\colon k\in\Gamma _0,k_0=1,2,3}\\
  &\quad\ctimes\sum\brac{\frac{1}{k_0}\,\sqbrac{a(k)+a(k)^*}\colon k\in\Gamma _0,k_0=1,2,3}
\end{align*}
Acting on $\ket{pq}$, the only terms that may give nonzero inner products are
\begin{align*}
\sum&\brac{\frac{1}{k_0^2}\,e^{i\pi k_0/2}\colon k\in\Gamma _0,k_0=1,2,3}\ket{pq}\\
  &=\paren{6e^{i\pi /2}+\frac{6}{2^2}\,e^{i\pi}+\frac{30}{3^2}\,e^{i3\pi (2)}}\ket{pq}=c_1\ket{pq}
\end{align*}
where $c_1=-1.50+2.67i$. The next term is
\begin{align*}
\sigma&\paren{(1,\bfe ),3}\sigma (0,3)\\
  &=\sum\brac{\frac{1}{k_0}\,\sqbrac{a(k)e^{i\pi (k_0-k_1)/2}+a(k)^*e^{-i\pi (k_1-k_0)/2}}\colon k\in\Gamma _0,k_0=1,2,3}\\
  &\qquad\ctimes\sum\brac{\frac{1}{k_0}\,\sqbrac{a(k)+a(k)^*}\colon k\in\Gamma _0,k_0=1,2,3}
\end{align*}
Acting on $\ket{pq}$, the only terms that may give nonzero inner products are
\begin{align*}
\sum&\brac{\frac{1}{k_0^2}\,e^{i\pi (k_0-k_1)/2}\colon k\in\Gamma _0,k_0=1,2,3}\ket{pq}\\
  &=\paren{4e^{i\pi /2}+\frac{1}{2}\,e^{i\pi}-\frac{4}{9}\,e^{i 3\pi /2}}\ket{pq}=c_2\ket{pq}
\end{align*}
where $c_2=-0.50+4.44i$. Since the other terms contribute this same coefficient $c_2$, the nonzero part of $H(1,3)H(0,3)$ is
\begin{equation}         
\label{eq54}
H_{1,0}=\frac{1}{7}\,\sqbrac{c_1\phi\paren{(1,\bfzero ),3}^2+c_2\phi\paren{(1,\bfe ),3}^2+\cdots +c_2\phi\paren{(2,-\bfg ),3}^2}\phi (0,3)^2
\end{equation}

We now consider the terms of $H_{1,0}$. We have that
\begin{align}         
\label{eq55}
\phi&\paren{(1,\bfe ),3}^2\phi (0,3)^2\notag\\
  &=\sqbrac{\sum\brac{\frac{1}{p_0}\,\sqbrac{a(p)e^{i\pi p_0/2}+a(p)^*e^{-i\pi p_0/2}}\colon p\in\Gamma _m,p_0=1,2,3}}^2\notag\\
  &\qquad\ctimes\sqbrac{\sum\brac{\frac{1}{p_0}\,\sqbrac{a(p)+a(p)^*}\colon p\in\Gamma _m,p_0=1,2,3}}^2
\end{align}
There are six possibilities for the interaction $\ket{pq}\to\ket{p'q'}$. These can be classified according to Feynman diagrams but we shall just use the following symbolic notation:
{\obeylines
(Case 1)\enspace$\ket{pq}\to\ket{0}\to\ket{p'q'}$
(Case 2)\enspace$\ket{pq}\to\ket{p'q}\to\ket{p'q'}$
(Case 3)\enspace$\ket{pq}\to\ket{q'q}\to\ket{p'q'}$
(Case 4)\enspace$\ket{pq}\to\ket{pp'}\to\ket{p'q'}$
(Case 5)\enspace$\ket{pq}\to\ket{pq'}\to\ket{p'q'}$
(Case 6)\enspace$\ket{pq}\to\ket{pqp'q'}\to\ket{p'q'}$
}\bigskip

Letting $\beta =1/p_0q_0p'_0q'_0$, the term $\bra{p'q'}\phi\paren{(1,\bfzero ),3}^2\phi (0,3)^2\ket{pq}$ contributes the coefficient $\beta e^{i\pi\alpha /2}$ where
\begin{equation*}
\alpha =-p'_0-q'_0,q_0-q'_0,q_0-p'_0,p_0-q'_0,p_0-p'_0,p_0+q_0
\end{equation*}
in the six cases. Adding these six coefficients and applying conservation of energy-momentum gives the following contributions of \eqref{eq55}
\begin{equation*}
d_1=2\beta\sqbrac{\cos\frac{\pi}{2}\,(p_0+q_0)+\cos\frac{\pi}{2}\,(q_0-q'_0)+\cos\frac{1}{2}\,(q_0-p'_0)}
\end{equation*}

The next term in \eqref{eq54} is
\begin{align*}
\phi&\paren{(1,\bfe ),3}^2\phi (0,3)^2\\
&=\sqbrac{\sum\brac{\frac{1}{p_0}\,\sqbrac{a(p)e^{i\pi (p_0-p_1)/2}+a(p)^*e^{-i\pi (p_0-p_1)/2}}\colon p\in\Gamma _m,p_0=1,2,3}}^2\\
&\qquad\ctimes\sqbrac{\sum\brac{\frac{1}{p_0}\,\sqbrac{a(p)+a(p)^*}\colon p\in\Gamma _m,p_0=1,2,3}}^2
\end{align*}
Again, we have six cases for $\beta e^{i\pi\alpha /2}$
\begin{align*}
\alpha&=p'_1+q'_1-p'_0-q'_0,q_0-q'_0-q_1+q'_1,q_0-p'_0-q_1+p'_1,\\
  &\quad p_0-q'_0-p_1+q'_1,p_0-p'_0-p_1+p'_1,p_0+q_0-p_1-q_1
\end{align*}
The sum of these coefficients gives
\begin{align*}
d_2&=2\beta\left[\cos\frac{\pi}{2}\,(p_0+q_0-p_1-q_1)+\cos\frac{\pi}{2}\,(q_0-q'_0-q_1+q'_1)\right.\\
  &\qquad\left. +\cos\frac{\pi}{2}\,(q_0-p'_0-q_1+p'_1)\right]
\end{align*}

The next term in \eqref{eq54} has the form $\phi\paren{(1,-\bfe ),3}^2\phi (0,3)^2$. The sum of the corresponding coefficients $d_3$ will be the same as $d_2$ except the $p_1,q_1,p'_1,q'_1$ terms will be the negative of those in $d_2$. We then have that 
\begin{align*}
d_2+d_3&=4\beta\left[\cos\frac{\pi}{2}\,(p_0+q_0)\cos\frac{\pi}{2}\,(p_1+q_1)+\cos\frac{\pi}{2}\,(q_0-q'_0)\cos\frac{\pi}{2}\,(q_1-q'_1)\right.\\
  &\qquad\left. +\cos\frac{\pi}{2}\,(q_0-p'_0)\cos\frac{\pi}{2}\,(q_1-p'_1)\right]
\end{align*}
The other terms will be similar, so adding all the terms gives
\begin{align}         
\label{eq56}
\bra{p'q'}S(2,3)\ket{pq}&=\frac{1}{7}\,\sqbrac{c_1d_1+c_2(d_2+d_3+d_4+d_5+d_6)}\notag\\
  &=-\frac{2\beta}{7}\left\lbrace\cos\frac{\pi}{2}\,(p_0+q_0)\,\sqbrac{c_1+2c_2\sum _{j=1}^3\cos\frac{\pi}{2}\,(p_j+q_j)}\right.\notag\\
  &\qquad +\cos\frac{\pi}{2}\,(q_0-q'_0)\sqbrac{c_1+2c_2\sum _{j=1}^3\cos\frac{\pi}{2}\,(q_j-q'_j)}\\
  &\qquad \left.+\cos\frac{\pi}{2}\,(q_0-p'_0)\sqbrac{c_1+2c_2\sum _{j=1}^3\cos\frac{\pi}{2}\,(p'_j-q_j)}\right\rbrace\notag
\end{align}
We now consider our specific example, $p=(2,0,-1,0)$, $q=(3,-1,2,1)$, $p'=(2,0,0,-1)$, $q'=(3,-1,1,2)$. We then have $\beta =1/36$ and applying \eqref{eq56} gives the amplitude
\begin{align}         
\label{eq57}
A_{2,3}\paren{\ket{pq}\to\ket{p'q'}}&=\bra{p'q'}S(2,3)\ket{pq}=\frac{-1}{126}\,(c_1+2c_2)\notag\\
  &=0.019841-0.09456i
\end{align}
We get the same result for $A_{2,3}\paren{\ket{pq}\to\ket{p''q''}}$.

We still have the interaction $\ket{pq}\to\ket{pq}$ which is harder to compute. In this situation we have the following cases.
{\obeylines
(Case 1')\enspace$\ket{pq}\to\ket{0}\to\ket{p'q'}$
(Case 2')\enspace$\ket{pq}\to\ket{rq}\to\ket{p'q'}$
(Case 3')\enspace$\ket{pq}\to\ket{ps}\to\ket{p'q'}$
(Case 4')\enspace$\ket{pq}\to\ket{pqrs}\to\ket{p'q'}$
}\bigskip

\noindent where $r,s\in\Gamma _m$, $r_0,s_0=1,2,3$. It follows from the $m=\sqrt{3}$ examples in Section~2 that the different possible $r$ and $s$ values are in the set
\begin{align*}
S&=\left\lbrace (2,\pm 1,0,0),(2,0,\pm 1,0),(2,0,0,\pm 1),(3,\pm 2,\pm 1,\pm 1),\right.\\
  &\quad\left. (3,\pm 1,\pm 2,\pm 1),(3,\pm 1,\pm 1,\pm 2)\right\rbrace
\end{align*}

As before, consider the term $\phi\paren{(1,\bfzero ),3}^2\phi (0,3)^2$. Case~1' contributes the coefficient
\begin{equation*}
\beta e^{-i\pi (p_0+q_0)/2}=-i\beta
\end{equation*}
Case~2' contributes the coefficient
\begin{equation*}
\sum\brac{\frac{1}{p_0^2r_0^2}\,e^{i\pi (r_0-p_0)/2}\colon r\in S}=\frac{6}{16}-\frac{24}{81}\,i=0.375-0.667i
\end{equation*}
Case~3' contributes the coefficient
\begin{equation*}
\sum\brac{\frac{1}{q_0^2s_0^2}\,e^{i\pi (s_0-q_0)/2}\colon s\in S}=\frac{-6}{36}\,i+\frac{24}{81}=0.296-0.167i
\end{equation*}
Case~4' contributes the coefficient
\begin{equation*}
\sum\brac{\frac{1}{r_0^2s_0^2}\,e^{i\pi (r_0+s_0)/2}\colon r,s\in S}=-4.783+8i
\end{equation*}
Adding these together gives
\begin{equation*}
d'_1=4.1111+7.1389i
\end{equation*}

For the term $\phi\paren{(1,\bfe ),3}^2\phi (0,3)^2$, Case~1' contributes the coefficient
\begin{equation*}
\beta e^{-i\pi (p_0+q_0-p_1-q_1)/2}=-\beta
\end{equation*}
Case~2' contributes the coefficient
\begin{align*}
\sum\brac{\frac{1}{p_0^2r_0^2}\,e^{i\pi (r_0-p_0-r_1+p_1)/2}\colon r\in S}&=-\sum\sqbrac{\frac{1}{p_0^2r_0^2}\,e^{i\pi (r_0-r_1)/2}\colon r\in S}\\
  &=0.250-0.222i
\end{align*}
Case~3' contributes the coefficient
\begin{align*}
\sum\brac{\frac{1}{q_0^2s_0^2}\,e^{i\pi (s_0-q_0-s_1+q_1)/2}\colon s\in S}&=-\sum\sqbrac{\frac{1}{q_0^2s_0^2}\,e^{i\pi (s_0-s_1)/2}\colon s\in S}\\
  &=-0.111+0.0988i
\end{align*}
Case~4' contributes the coefficient
\begin{equation*}
\sum\brac{\frac{1}{r_0^2s_0^2}\,e^{i\pi (r_0-s_0-r_1-s_1)/2}\colon r,s\in S}=0.9474-1.333i
\end{equation*}
Adding these last three terms together gives
\begin{equation*}
d'_2=1.0864-1.4568i
\end{equation*}

The next term is $\phi\paren{(1,-\bfe ),3}^2\phi (0,3)^2$. Case~1' contributes the coefficient
\begin{equation*}
\beta e^{-i\pi (p_0+q_0+p_1+q_1)/2}=\beta
\end{equation*}
Now Cases~2',~3' and 4' give the same coefficients as for the previous terms. Similar results apply for the remaining four terms. Applying \eqref{eq51} we obtain the amplitude
\begin{align}         
\label{eq58}
A_{2,3}\paren{\ket{pq}\to\ket{pq}}&=\bra{p'q'}S(2,3)\ket{pq}=1+\frac{1}{7}\,(c_1d'_1+6c_2d'_2)\notag\\
  &=-2.2455-1.667i
\end{align}

We now apply reconditioning to obtain probabilities at the $(2,3)$ perturbation level. By \eqref{eq57} and \eqref{eq58} we have the normalization constant
\begin{equation*}  
N(2,3)^2=2\ab{0.019841-0.094356i}^2=\ab{-2.2455-1.6671i}^2=7.8401
\end{equation*}
The probabilities become
\begin{align*}
P_{2,3}\paren{\ket{pq}\to\ket{p'q'}}&=P_{2,3}\paren{\ket{pq}\to\ket{p''q''}}\\
  &=\frac{\ab{A_{2,3}\paren{\ket{pq}\to\ket{p'q'}}}^2}{N(2,3)^2}=0.0011858\\
  P_{2,3}\paren{\ket{pq}\to\ket{pq}}&=\frac{\ab{A_{2,3}\paren{\ket{pq}\to\ket{pq}}}^2}{N(2,3)^2}=0.99763
\end{align*}

We conclude that at times $x_0=0$ and $x_0=1$, nontrivial scattering does not take place. Nontrivial scattering begins at $x_0=2$ when it is very small. We conjecture that the amount of scattering will increase with time. In particular, we conjecture that
\begin{equation*}  
\lim _{x_o,r\to\infty}P_{x_0,r}\paren{\ket{pq}\to\ket{pq}}=0
\end{equation*}
and that
\begin{equation*}  
\lim _{x_o,r\to\infty}P_{x_0,r}\paren{\ket{pq}\to\ket{p'q'}}=\lim _{x_0,r\to\infty}P_{x_0,r}\paren{\ket{pq}\to\ket{p''q''}}=1/2
\end{equation*}
Proving this conjecture may be quite difficult and providing plausibility would probably require computer aided numerical calculations.

\section{A Test for Discreteness} 
We have seen in Table~1 that for each energy (frequency) there are finitely many momentum directions for a photon. The same conclusion holds for particles with a specific total energy and mass. In this way, photons with a specific energy can only propagate in a finite number of directions. If spacetime is discrete, then probably the simplest mode describing it is a 4-dimensional cubic lattice. The edges of this lattice presumably would have Planck length of about $10^{-33}$cm. It is unlikely that we will ever develop instruments accurate enough to measure anywhere near such lengths or the corresponding Planck times of about $10^{-43}$sec. In this respect, it appears impossible to directly observe a granular, discrete structure for spacetime. However, if we consider the dual energy-momentum space, a different situation appears. Unlike lengths, angles expand with distance from an origin and instead of investigating the very small, we should study the very large.

Consider photons with a specific frequency propagating from a distant star which can be considered to be essentially a point source. The photons may be able to move in millions of directions from their point source, but after a thousand light years their angular spread will be quite large. In a sense, the small granular structure of space-time is greatly amplified. It is then possible that a human eye or a telescope on earth will be situated in a ``shadow'' region that the photons do not reach.

Astronomers tell us that stars appear to twinkle because their light is bent or refracted when it passes through turbulence in the earth's atmosphere. But there could be a secondary reason for this twinkling. As the earth spins and orbits the sun, it can move in and out of photon shadows causing a twinkling effect. Suppose we consider a telescope like the Hubble telescope that is in an earth orbit. Such a telescope is outside the earth's atmosphere so by conventional wisdom, stars it observes should not twinkle. I suggest that the effect may be small, but if careful measurements are made, then the stars will still twinkle! This would be a test for the discreteness of spacetime.

This may also give an explanation for dark energy and matter. With this reasoning there actually is no dark energy and matter. It is ordinary energy and matter that we can not detect. We are in the shadow regions of many photons so we do not observe them. In a similar way, there are many stars or even galaxies that we cannot see or that have very weak images because we are in their shadow regions.

\end{document}